# Post-synthesis tuning of dielectric constant via ferroelectric domain wall engineering

L. Zhou[1,2], L. Puntigam[1], P. Lunkenheimer[1], E. Bourret[3], Z. Yan[3,4], I. Kézsmárki[1],
D. Meier[5], S. Krohns[1], J. Schultheiß[5], and D. M. Evans[1,6a)]

[1]*Experimental Physics V, Center for Electronic Correlations and Magnetism, Institute of Physics, University of Augsburg, 86159 Augsburg, Germany*

[2]*School of Physics, University College Dublin, Dublin D04 V1W8, Ireland*

[3]*Materials Sciences Division, Lawrence Berkeley National Laboratory, Berkeley, California 94720, USA*

[4]*Department of Physics, ETH Zurich, 8093 Zürich, Switzerland* [2]*Department of Materials Science and Engineering,*

[5]*Norwegian University of Science and Technology (NTNU), 7043 Trondheim, Norway*

[6]*Department of Physics, University of Warwick, Coventry CV4 7AL, UK*

Email: Donald.Evans@warwick.ac.uk

**Abstract**

A promising mechanism for achieving colossal dielectric constants is to use insulating internal barrier layers, which typically form during synthesis and then remain in the material. It has recently been shown that insulating domain walls in ferroelectrics can act as such barriers. One advantage domain walls have, in comparison to stationary interfaces, is that they can be moved, offering the potential of post-synthesis control of the dielectric constant. However, to date, direct imaging of how changes in domain wall pattern cause a change in dielectric constant within a single sample has not been realized. In this work, we demonstrate that changing the domain wall density allows the engineering of the dielectric constant in hexagonal-ErMnO$_3$ single crystals. The changes of the domain wall density are quantified via microscopy techniques, while the dielectric constant is determined via macroscopic dielectric spectroscopy measurements. The observed changes in the dielectric constant are quantitatively consistent with the observed variation in domain wall density, implying that the insulating domain walls behave as 'ideal' capacitors connected in series. Our approach to engineer the domain wall density can be readily extended to other control methods, e.g., electric fields or mechanical stresses, providing a novel degree of flexibility to *in-situ* tune the dielectric constant.

## I. Introduction

Materials with high dielectric constants are of considerable importance for applications, playing a key role, e.g., for short-term energy storage in capacitors [1–3], and they ensure the performance of today's smart phones, portable computers, and electronic vehicles. For this, a high dielectric constant and low loss tangent are important criteria [4]. Materials with very high dielectric permittivity ($\varepsilon' > 1000$) are often referred to as having "colossal dielectric constants" (CDC) [5,6], offering performance parameters of great technological interest. There are two typical approaches to achieve such "colossal" values: i) choose a material which naturally exhibits such traits using, e.g., the CDC that arises across a ferroelectric transition [7]; or ii) engineer insulating layers within a material to act as barrier layer capacitors (BLC) [5,8]. What both of these approaches have in common is the colossal values that are achieved during material synthesis and then "locked-in" to the material. An example of approach i) is Pb($Zr_x$,$Ti_{1-x}$)$O_3$ (PZT), which displays both a CDC and low loss tangent [9]. The room temperature dielectric constant of PZT can exceed $10^3$ with a loss tangent of tan $\delta$ < $10^{-2}$. A common example of ii) are BLCs observed in ceramics, where they form at insulating grain boundaries, and at metal-semiconductor interfaces due to the formation of a Schottky barrier [5,6,10,11]. Such barrier layers are associated with a step-like increase in $\varepsilon'$ and a peak in $\varepsilon''$, representing a "Debye-like", so-called Maxwell-Wagner (MW) relaxation process [5,12–14]. For a more detailed discussion, the interested reader is referred to, for example, the review of Lunkenheimer *et al.* [5].

The concept of using internal barrier layers at grain boundaries [8,12] has recently been expanded towards ferroelectric domain walls (DWs) [15,16]. Such DWs naturally arise as a result of the spontaneous polarization domains, and their density can readily be engineered via, e.g., electric fields, mechanical pressure, microstructural levers, or annealing, which gives additional opportunities for tailoring dielectric properties [17]. Particularly intriguing in this context are improper ferroelectrics, where the polarization is not the primary symmetry breaking order parameter, as it has been discovered that these systems can develop stable insulating DWs [18,19] that can act as BLCs [15,16]. Further, the DWs in improper ferroelectrics are particularly stable to electric fields, compared to proper ferroelectrics, because the electric field is not the conjugate field to the primary order parameter [20,21]. To date, however, the concept has only been explored at a proof-of-concept level [15,16], whereas direct imaging of the domain wall density, combining piezoresponse force microscopy (PFM) and scanning electron microscopy (SEM), together with a corresponding quantitative derivation of the changes in the macroscopic dielectric constant originating from changes in DW density in a single sample is missing.

The family of hexagonal manganites has a room temperature $P6_3cm$ symmetry arising from a high temperature $P6_3/mmc$ phase transition [22]. It was among the first material families to show stable charged DWs with unusual electronic properties [18,23,24]. In this work, we choose single-crystalline hexagonal $ErMnO_3$ as a template material as it is a well-studied improper ferroelectric (P ≈ 5.5 μC/cm$^2$ and $T_C$ ≈ 1150 °C), with stable charge DWs [25,26]. We show that post-synthesis engineering of the DW density in $ErMnO_3$ changes the material's dielectric constant. We quantify the DW density and variations in the dielectric constant by applying different microscopy techniques, as well as dielectric measurements recorded before and after thermal annealing. We observe a direct correlation between changes in the density of (insulating) DWs and the macroscopic dielectric response, establishing DW engineering as a powerful tool for achieving tunable and colossal dielectric constants post-synthesis.

## II. Experimental methods

The ErMnO$_3$ crystal used in this study is cut from the same single crystal used by Puntigam *et al*. in ref. [15], and was grown by the pressurized floating zone method [27]. The sample was measured as prepared and after subjecting it to thermal treatment as indicated in Fig. 1(a). For the pre-annealing measurements the sample had an area of 2.18 mm$^2$ and thickness of 1.74 mm. During annealing some of the sample broke off and so the sample had an area of 1.72 mm$^2$ and a thickness of 1.72 mm for the post annealing measurements.

The dielectric data was collected with an Alpha-a high-performance frequency analyzer from Novocontrol Technologies (frequency range: 3 µHz – 10 MHz) applying an ac voltage of 1 V. The recorded data are conductance, $G'$, and the real part of the complex capacitance, $C'$. On the pre-annealing state, the measurements were conducted in a Janis research SHI-950 refrigerator system cold head cryostat, and the post-annealing state was measured in a closed cycle refrigerator (Cryodyne Refrigeration System Model 22 from CTI-Cryogenics). For both measurements the sample was prepared in a parallel-plate capacitor geometry with top and bottom electrodes made from silver paste (Leitsilber 200N manufactured by Hans Wolbrig).

The SEM images were collected with a ZEISS Crossbeam 550 with ZEISS Gemini 2 electron optics, using the in-lens detector, with an electron beam of 500 pA and 5 kV (Fig. 1(b)) and 200 pA and 5 kV (Fig. 1(c)). The PFM experiments were carried out on a Bruker Dimension Icon AFM, using n-doped silicon tips coated with Pt-Ir (SCM-PIT-V2, Bruker). The PFM was collected with an amplitude of 10 V and resonance frequencies of 71.01 kHz and 69.47 kHz, in Fig. 1(b) and Fig. 1(c), respectively.

## III. Control of microstructure

To quantify the DW density in our ErMnO$_3$ crystal, we image the ferroelectric domain structure using PFM and SEM. The initial domain structure is presented in Fig. 1(b). The gold-brown side panel of Fig. 1(b) shows the in-plane piezoresponse of ErMnO$_3$, revealing the characteristic domain pattern with six-fold vertices* as explained elsewhere [22,28,29]. The PFM measurement is complemented by the top SEM image (black and white) of the polar surface, corroborating the formation of an isotropic domain structure (see, e.g., refs. [30] and [31] for details concerning SEM contrast formation in hexagonal manganites). Based on the imaging experiments, the vertex density can be estimated by counting the number of vertexes present and dividing that value by the area of the image. This gives a vertex density of approximately 0.5 vertices / $\mu m^2$ in the as-grown state.

---

*In this work we interpret the word "vortex" to represent a continuous rotation, e.g., the continuous rotation of the structural order parameter in the hexagonal manganites, while we use the word "vertex" for the meeting point of multiple objects, e.g., the six-fold meeting point of different ferroelectric domains in the hexagonal manganites. Because this work focuses on the polarization properties, we use the word vertex to refer to these meeting points.

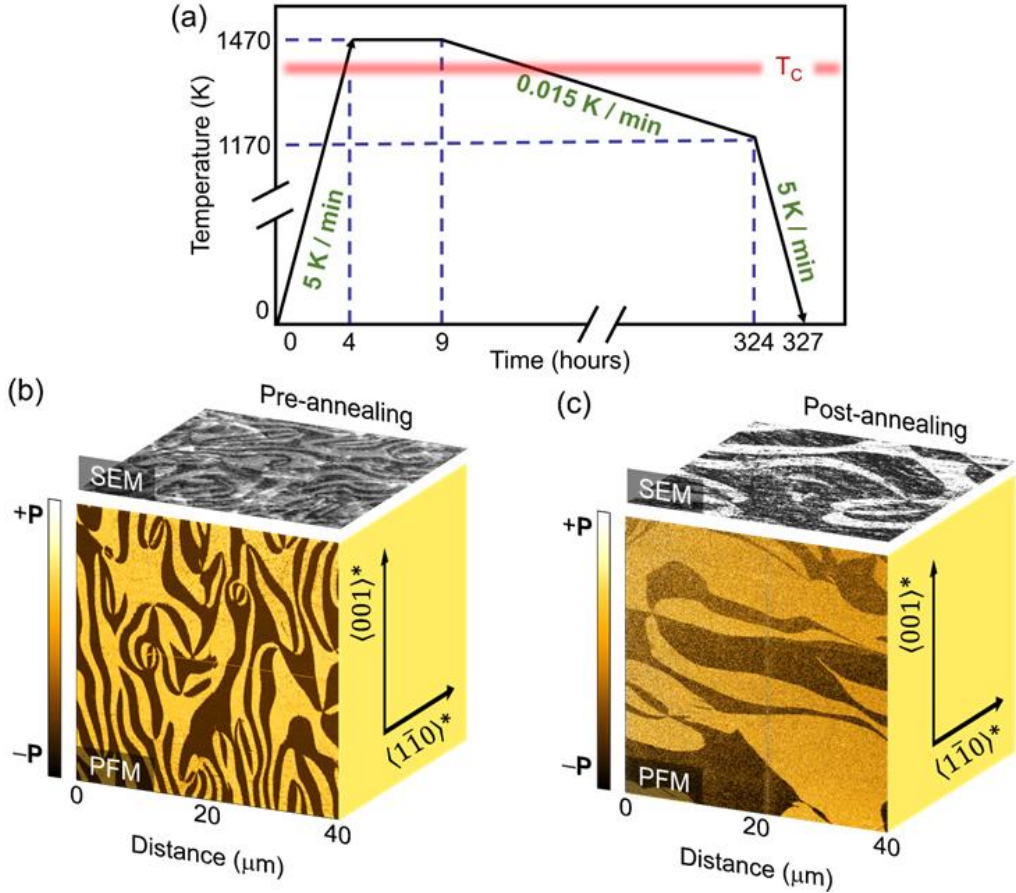

FIG 1. Ferroelectric domain structure before, **a**, and after, **c**, thermal treatment. Both frames show representative images of the ferroelectric domain structure recorded on the non-polar surface with PFM (in-plane contrast) and the polar surface with SEM. **b** shows the protocol used for the thermal treatment, where the Curie temperature ($T_C$) is represented by the red line. The crystallographic directions are in a pseudocubic notation, with the polar direction perpendicular to the black and white SEM image surface.

By applying the thermal treatment illustrated in Fig. 1(a), the vertex density in ErMnO$_3$ can be controlled [22,32,33], providing a viable handle for a reduction of the number of DWs in the system. This is confirmed by the PFM and SEM images in Fig. 1(c), which are recorded after thermal annealing (with a cooling rate of 0.015 K/min) through the transition region around the Curie temperature ($T_C \approx 1150$ °C) [32]. We note that after annealing, the sample was re-polished to ensure a flat surface for the PFM and SEM imaging of bulk properties following the procedure outlined in ref. [15]. Both the SEM and PFM images reveal a significantly reduced density of vertices, i.e., around 0.1 / $\mu m^2$, and consequently an increased domain size. As each vertex should connect six DWs, this observation shows that the number of DWs, including the insulating (head-to-head [18]) DWs, is reduced by nearly an order of magnitude.

## IV. Dielectric properties

After demonstrating that the vertex density, hence domain wall density, is reduced by the applied thermal annealing procedure, consistent with the literature, we next turn to the bulk dielectric properties [32]. The real part of the dielectric constant $\varepsilon'$ and the loss tangent tan $\delta$ are recorded for the pre-annealing state, i.e., for the domain state seen in Fig. 1(b), and are given in Figs. 2(a), and (b), respectively. $\varepsilon'(T)$ exhibits a rapid increase with increasing temperature before plateauing off. This increase shifts to higher temperature for higher frequencies, moving outside our measurement range for frequencies above ~ 12 kHz. After thermal treatment (leading to the domain state in Fig. 1(c)), marked differences in both the real part and tan $\delta$ were observed, as shown in Figs. 2(c) and (d), respectively. Notably, the rapid increase now occurs approximately 50 K lower in temperature, and a second step-like increase is now evident, as seen, e.g., in the 1 Hz dataset of Fig. 2(c).

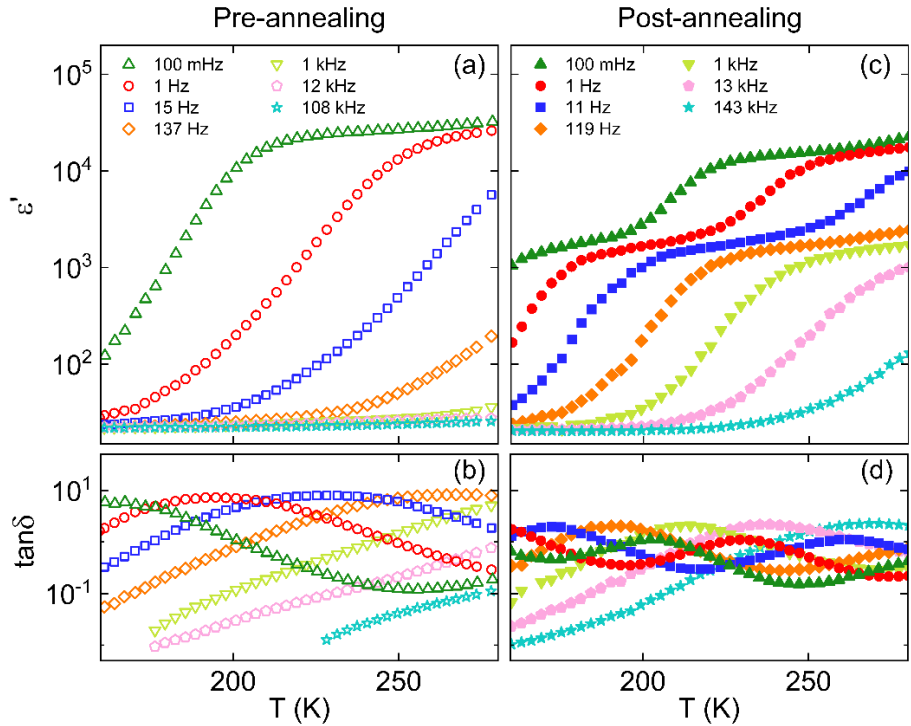

FIG 2. Effect of changing DW density on the dielectric response. Panel (a) shows the real part of the dielectric constant as a function of temperature, while panel (b) shows the loss tangent, tan δ, for representative frequencies. Both are taken on the pre-annealing state. Panels (c) and (d) display the real part of the dielectric constant and the loss tangent, respectively, taken after thermal treatment was used to alter the domain structure, i.e. for the post-annealing state.

Additionally, Figs. 2(b) and (d) reveal that the step-like increases correspond to peaks in tan δ, indicating that they originate from relaxation processes. Relaxation-like features, as in Figs. 2(a)-(d), are common in oxide materials and often originate from internal BLCs (IBLCs) and/or surface BLCs (SBLCs) [5,6]. IBLCs arise from MW polarisation effects at interfaces within the material, such as DWs, whereas SBLCs often result from the formation of Schottky barriers at the sample-electrode interfaces and the associated depletion zone acting as a thin insulating layer [5,6].

For a more detailed analysis of the change in dielectric properties, in Fig. 3 we present the spectra at four representative temperatures. Figure 3(a) shows the frequency-dependent $\varepsilon'$ values for 150 K, 190 K, 230 K, and 270 K. In addition, Fig. 3(b) provides frequency-dependent conductivity data, $\sigma'(\nu)$, at the same temperatures. Using the 270 K data of Fig. 3(a) as a representative example, at high frequencies ($\nu \gtrsim 10^3$ Hz) $\varepsilon'$ flattens out to nearly a constant value, consistent with data reported in the literature [15,16]. At lower frequencies ($\nu \lesssim 10^3$ Hz), there is a relaxation-like feature in $\varepsilon'$, leading to a step-like increase by about three decades to colossal values of $\varepsilon'$ of the order of $10^4$.

Figure 3(c) and (d) show $\varepsilon'$ and $\sigma'$ measured after thermal treatment, i.e., for the sample with a reduced number of DWs. As before, we use the 270 K data of Fig. 3(c) as a representative example of the post-annealing properties. While the dielectric constant approached at high frequencies is the same as for the untreated sample, at lower frequencies, there are now two distinct relaxations instead of the single process observed pre-annealing (Fig. 3(a)). For instance, at 270 K in Fig. 3(c), there is a step-like increase in $\varepsilon'$ from ~21 to ~2500, for $10^3$ Hz $\lesssim \nu \lesssim 10^5$ Hz, followed by a second step-like increase in $\varepsilon'$ from ~2500 to ~$2\times10^4$ for $\nu \lesssim 100$ Hz. A similar succession of relaxation features is also observed in the conductivity data in Fig. 3**d**. Here, one should note that the dielectric loss $\varepsilon''$ is related to $\sigma'$ via $\varepsilon'' \propto \sigma'/\nu$. Thus, the observed steps in $\sigma'(\nu)$ (Figs. 3(b),(d)) correspond to peaks in $\varepsilon''(\nu)$, shifting with temperature, which is the typical signatures of a relaxation process. Before thermal treatment (Fig. 3(b)), the 270 K curve shows a single plateau between ~ 10 Hz and ~ $10^5$ Hz at $\sigma' \sim 8\times10^{-8}$ (Ωcm)$^{-1}$ and a step-like decrease for $\nu \lesssim 1$ Hz. After thermal treatment (Fig. 3(c)) two plateaus are observed with a value of $\sigma' \sim 1.5\times10^{-5}$ (Ωcm)$^{-1}$ at frequencies $\gtrsim 10^5$ Hz. Both before and after treatment, we observe an increase of $\sigma'(\nu)$, superimposed to the high-frequency plateaus, which can be ascribed to a so-called "universal dielectric response" (UDR), where $\sigma'$ follows a power law behavior [34]. It is commonly ascribed to hopping charge transport [5,35] and most clearly revealed at 150 K, in the post-annealing state (Fig. 3(d)), where the relaxation-related step has shifted out of the frequency window. Notably, other studies on the closely related hexagonal manganite YMnO$_3$ showed qualitatively similar behavior [16], including the occurrence of two well-resolved MW relaxations with a reduced DW density (0.04 vertices / µm$^2$), whereas only a single clear relaxation step was observed in a different YMnO$_3$ sample with 1.5 vortices / µm$^2$. The key observation from Fig. 3 is that the density of DWs correlates with the dielectric properties of the same ferroelectric hexagonal manganites sample and, hence, can be used to tune them.

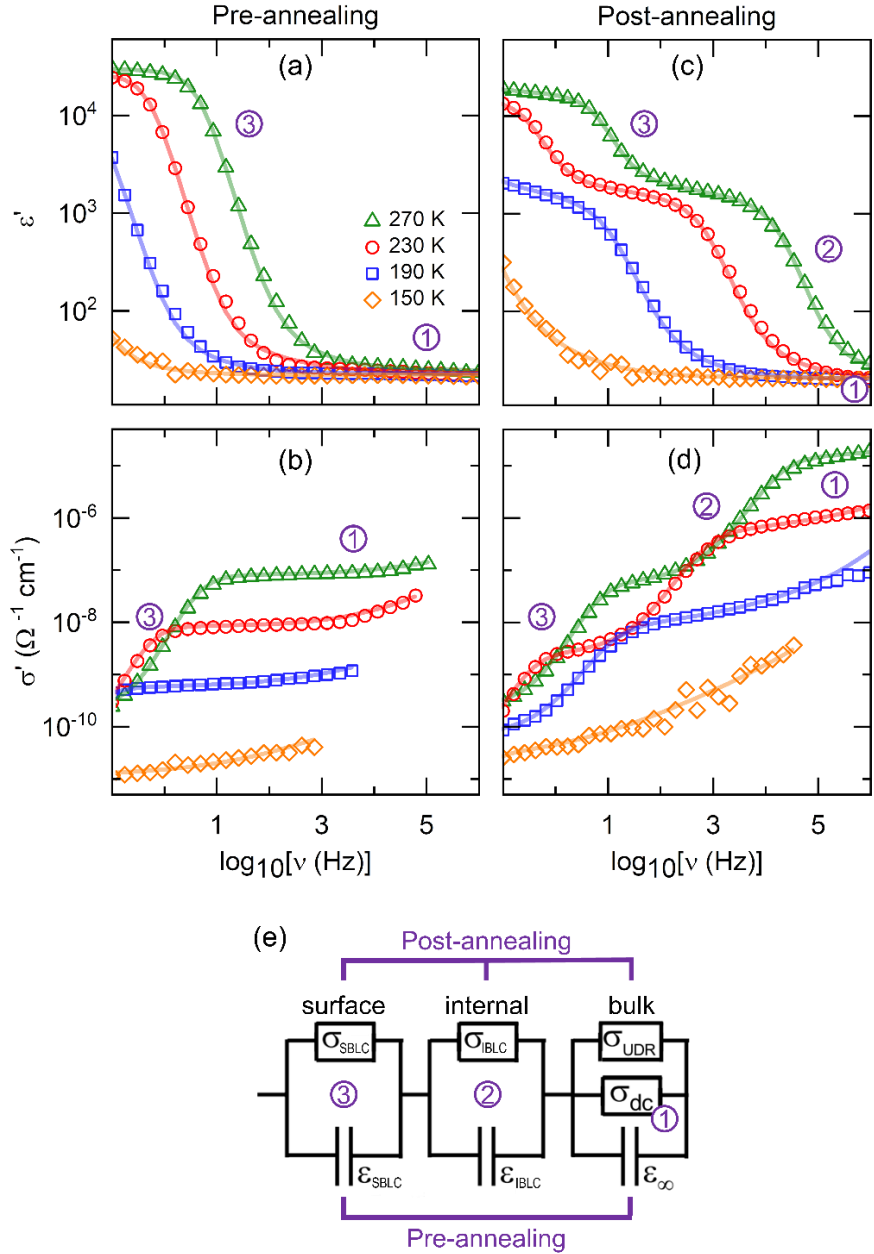

FIG 3. Frequency dependent dielectric permittivity and conductivity taken of the pre- and post-annealing, (a) and (b) and (c) and (d), respectively. (e) gives the equivalent circuit model used to fit the data[5]. Fits are given by solid lines in all panels, with fit parameters provided in Table I. In all panels the green triangles are collected at 270 K, the red circles at 230 K, the

## V. Dielectric analysis

In order to disentangle the different contributions to the dielectric response, we use an equivalent-circuit analysis to fit the frequency-dependent data. Following the established procedure in literature [5,6,16], individual parallel RC elements, connected in series to each other, are used to represent the different contributions: the surface barrier layers, internal barrier layers, and the bulk. This is pictorially represented by Fig. 3(e). The RC element for the bulk has an additional frequency-dependent impedance connected in parallel to represent the contribution of the UDR. A detailed description of this approach is provided in ref. [5], and examples of its implementation in this material class can be found in refs. [15,16,36]. The fits are shown by the solid lines in Figs. 3(a)-(d) provide a reasonable description of the experimental data. The corresponding fitting parameters given in Table I. The high-frequency plateau in $\sigma'(\nu)$ (denoted "1" in Figs. 3(b) and (d)) and the superimposed UDR represent the bulk response as here the interface contributions are shorted by their respective capacitances [5,6]. Correspondingly, at higher frequencies, $\varepsilon'(\nu)$ approaches its intrinsic value, which is of the order of 20 (Figs. 3(a) and (c)). In accordance with previous investigations of this and related materials [15,16], the high-frequency step in the post-annealing state (denoted "2") is attributed to IBLCs originating from insulating DWs, which leads to a MW relaxation. The second step at lower frequencies ("3") is ascribed to a MW relaxation caused by SBLCs, e.g., arising from the metal-semiconductor contacts [5,15,16]. In the pre-annealing state, only a single interface-related RC circuit is required to reproduce the experimental findings (denoted "3" in Fig 3(b)). This most likely implies that the MW relaxations from SBLC superimposes the IBLC relaxation. This changes after annealing: Comparing the high-frequency plateaus of $\sigma'(\nu)$ in Figs. 3(b) and (d) reveals that the bulk conductivity strongly increases upon thermal treatment. As the MW-relaxation time is proportional to the bulk resistivity [5,6], this moves the IBLC relaxation step to higher frequencies, allowing it to be distinguished from the dominating SBLC relaxation.

Table I. Fitting parameters for ε' and σ' used in Fig. 3. All values are rounded to 2 significant figures.

|  | T (K) | $\varepsilon_\infty$ ① | $\varepsilon'$ ② | $\varepsilon'$ ③ | $\sigma_{dc}^{bulk}$ [(Ω cm)$^{-1}$] ① | $\sigma_{dc}^{IBLC}$ [(Ω cm)$^{-1}$] ② | $\sigma_{dc}^{SBLC}$ [(Ω cm)$^{-1}$] ③ |
|---|---|---|---|---|---|---|---|
| Pre-annealing | 270 | 24 | - | 31000 | $7.6 \times 10^{-08}$ | - | $2.1 \times 10^{-10}$ |
|  | 230 | 23 | - | 28000 | $8.1 \times 10^{-09}$ | - | $5.3 \times 10^{-11}$ |
|  | 190 | 22 | - | 31000 | $5.4 \times 10^{-10}$ | - | $8.4 \times 10^{-12}$ |
| Post-annealing | 270 | 21 | 2500 | 20000 | $1.6 \times 10^{-05}$ | $8.0 \times 10^{-08}$ | $2.3 \times 10^{-10}$ |
|  | 230 | 21 | 2500 | 20000 | $5.7 \times 10^{-07}$ | $3.6 \times 10^{-09}$ | $2.7 \times 10^{-12}$ |
|  | 190 | 21 | 2500 | 20000 | $8.7 \times 10^{-09}$ | $9.2 \times 10^{-11}$ | $3.4 \times 10^{-14}$ |

Having experimentally established that changes in DW density impacts the dielectric spectra, next we aim to quantitatively estimate the influence of the DW density on the value of the colossal dielectric constants related to the MW relaxation process caused by IBLCs. To find this correlation, we make two reasonable assumptions [16]: (i) The interface resistance (the IBLC resistance in the present case) is much higher than the bulk resistance, i.e., we have insulating interfaces, acting as capacitors. (ii) The overall interface thickness is much smaller than the bulk thickness, i.e., in Fig. 3(e) the IBLC capacitance is much larger than the bulk capacitance. Compared to the intrinsic bulk dielectric constant $\varepsilon_b$, the colossal plateau of $\varepsilon'$ of an interface-generated MW relaxation ($\varepsilon_{col}$) then is generally enhanced by a factor that is given by the thickness ratio of the bulk and the interface [16,37]. This arises from the fact that the apparently colossal $\varepsilon_{col}$ is calculated using the bulk thickness instead of the several decades smaller interface thickness, the latter leading to large capacitance. When assuming that the intrinsic dielectric constant of the interface, $\varepsilon_i$, is the same as for the bulk, i.e., $\varepsilon_i = \varepsilon_b$, this leads to:

$$\varepsilon_{col} = \varepsilon_b \frac{d_b}{d_i}, \qquad (1)$$

where $d_b$ and $d_i$ are the thicknesses of the bulk sample and of the interface, respectively. In the present case, $\varepsilon_b$ is given by $\varepsilon_\infty$ of the MW relaxation, read off at high frequencies and low temperatures in Figs. 3(a) and (c), where the interface-related RC circuits are effectively short circuited, and the intrinsic response is detected. Then the above assumption $\varepsilon_i = \varepsilon_b$ (which in our case corresponds to $\varepsilon_{IBLC} = \varepsilon_\infty$, where $\varepsilon_{IBLC}$ is the intrinsic dielectric constant in the DWs) should be approximately fulfilled: It is reasonable to assume that the absolute value of $\varepsilon_b$ is dominated by induced dipoles, caused by the ionic and atomic polarizability. As the same ions and atoms exist both in the DWs and in the bulk, $\varepsilon_i$ (= $\varepsilon_{IBLC}$) and $\varepsilon_b$ (= $\varepsilon_\infty$) should be of similar order.

To utilize eq. (1), the next step is to find the total IBLC thickness, $d_i$, the sum of the thicknesses of all insulating DWs that the current has to cross. This is justified by the fact that a series connection of several identical capacitors is equivalent to a single capacitor with added-up thicknesses of the individual capacitors. Defining $d_{DW}$ as the thickness of a single DW and $n_{DW}$ as the number of insulating DWs per length scale crossed by the current, this leads to:

$$d_i = n_{DW}\, d_b\, d_{DW}. \qquad (2)$$

Putting eq. (2) into (1) and using $\varepsilon_b = \varepsilon_\infty$, we finally get:

$$\varepsilon_{col} = \frac{\varepsilon_\infty}{n_{DW}\, d_{DW}}. \qquad (3)$$

Thus, the presence of insulating DWs should lead to an effective enhancement of $\varepsilon'$ by a factor $f_{DW} = 1 / (n_{DW} d_{DW})$. To find $n_{DW}$ for our experimental data a line was drawn every 5 µm across the SEM image of the sample's domain structure, Figs. 1(b) and (c), and the number of intersections with DWs along the lines counted and averaged across all lines. As only some of the DWs are insulating in this system (see ref. [18] for a detailed discussion), the number of insulating DWs is estimated to be half of the total number of intersections. The quantity $d_{DW}$ corresponds to the thickness of the "electrically dressed walls", described by ref. [18] as the region in which the local conductance deviates from that of the bulk. Here we use a value of 150 nm to be consistent with ref. [15]. We also highlight, as mentioned in ref. [18], that this value is empirical and dependent on multiple factors including the measurement voltage and the angle of the polar discontinuity.

For the post-annealing state, our fits give $\varepsilon_\infty \sim 21$, and direct measurements of the domain structure as described above lead to $n_{DW} d_{DW} \sim 0.0116$ ($\pm 0.0038$) and $f \sim 86$. Using eq. (3), this allows us to derive a colossal static dielectric constant of the IBLC-related MW relaxation of $\varepsilon_{col} \sim 1810$ ($\pm 620$). This is in reasonable agreement with the value found from our fits to the spectra in Fig. 3, $\varepsilon_{col} = 2500$ ($\pm 200$). While it is not possible to experimentally derive $\varepsilon_{col}$ of the invisible IBLC-related MW-relaxation for the untreated sample, we can estimate its $\varepsilon_{col}$ value using the same formula, which gives $\varepsilon_{col} \sim 600$ ($\pm 80$). This smaller value of $\varepsilon_{col}$ reflects the larger number of DWs passed by the current in this sample, leading to higher total interface thickness. $\varepsilon_{col} \sim 600$ is much smaller than the colossal static dielectric constant ascribed to the SBLCs in this sample (about $3 \times 10^4$; cf. Fig. 3(a) and Table I), making it plausible that here the IBLC relaxation is superimposed by the dominant SBLC relaxation. The smaller expected value $\varepsilon_{col} \sim 600$ from IBLCs also confirms that the large, single MW relaxation, observed in the pre-annealing state, was correctly assigned to SBLCs.

Table II, A comparison of values from this work with literature.

| Sample | $\varepsilon_\infty$ | $n_{DW}$ (1/µm) | $n_{DW} d_{DW}$ (%) | estim. $\varepsilon_{col}$ | meas. $\varepsilon_{col}$ | calc. $n_{DW}$ (1/µm) |
|---|---|---|---|---|---|---|
| h-ErMnO$_3$ pre-annealing | 23 ($\pm 2$) | 0.257 ($\pm 0.025$) | 3.86 ($\pm 0.38$) | 600 ($\pm 80$) | - | - |
| h-ErMnO$_3$ post-annealing | 21 ($\pm 2$) | 0.077 ($\pm 0.025$) | 1.16 ($\pm 0.38$) | 1810 ($\pm 620$) | 2500 ($\pm 200$) | 0.056 ($\pm 0.007$) |
| h-ErMnO$_3$ [15] | 32 ($\pm 3$) | 0.50 ($\pm 0.05$) | 7.50 ($\pm 0.75$) | 430 ($\pm 60$) | 250 ($\pm 50$) | 0.85 ($\pm 0.19$) |
| h-(Er:Ca)MnO$_3$ [15] | 18 | 0.46 | 7 | 260 | 220 | 0.54 |
| h-YMnO$_3$ #1 [16] | 20 | 0.17 | 2.5 | 800 | 670 | 0.20 |

To highlight the key point that a change in the volume fraction of DWs allows post-synthesis engineering of the dielectric response, we plot the data of Table II in Fig. 4. The symbols show the measured data points and the trend line follows eq. (3). From Fig. 4 and eqs. (1) - (3), it is clear that the ideal structure to get a colossal dielectric constant would be a sample with a single, very thin insulating DW. This is readily explained if the individual walls behave as 'ideal' capacitors added in series, so every additional capacitor reduces the total value. For a single DW, eqs. (2) and (3), assuming an increase in the number of DWs for increasing sample thickness, are no longer valid and we have to revert to the more general eq. (1). It implies that the colossal $\varepsilon'$ then can be even further enhanced by using a thicker crystal. Note, the error bars represent the quantifiable errors, while the dominant error is expected to be the extent that surface images of the domain structure can be extrapolated into the bulk of the sample.

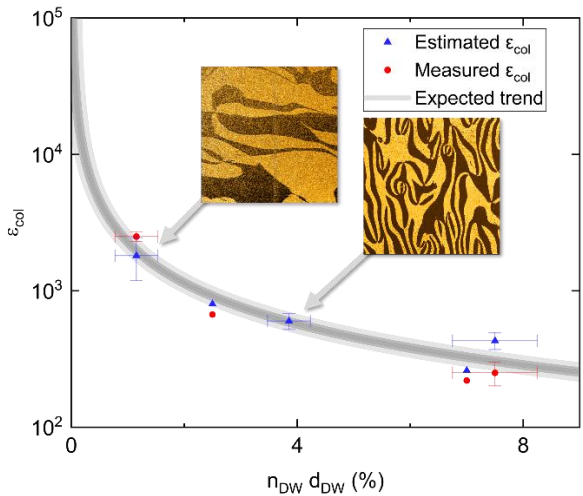

FIG 4. $\varepsilon_{col}$ contribution as a function of $n_{DW}\,d_{DW}$. The data points come from Table II, while a guide for the eyes is given in form of a line from eq. (3) assuming $\varepsilon_\infty = 22.8$. The inserts are the PFM images from Fig. 1, both are $40 \times 40$ µm.

## VI. Conclusion

We have demonstrated the ability to control the dielectric properties of ferroelectrics by adjusting the density of insulating DWs. In the case of ErMnO$_3$, it is correlated with the emergence of characteristic six-fold meeting points (vertex lines), which can readily be controlled via thermal treatment, providing a viable pathway for adjusting the dielectric response of the system. Our findings show a close, quantitative correlation between the DW density within the same samples and the resulting dielectric constant, which supports previous reports that use different densities of DWs across multiple samples. Consistent with previous studies, we have observed that the insulating DWs behave as ideal capacitors added in series at these voltages. These findings demonstrate the feasibility of post-synthesis engineering of dielectric responses in materials, with potential for expansion to other control methods such as electric fields and pressure. This approach offers a novel degree of flexibility and the possibility of dynamic real-time tuning of the dielectric constant.

## VII. Acknowledgments

The authors thank Dr. M. Altthaler for his skill and experience leading L.Z. in the collection of the SEM images. ChatGPT was used for proof reading and improving the flow of the text. Crystals were grown at the Lawrence Berkeley Laboratory supported by the U.S. Department of Energy, Office of Science, Basic Energy Sciences, Materials Sciences and Engineering Division (Contract No. DE-AC02-05-CH11231). L.P., P.L., and S.K. acknowledge the financial support of Deutsche Forschungsgemeinschaft Sachbeihilfe (DFG) via the Transregional Research Collaboration TRR 80 (Augsburg, Munich and Stuttgart; Project No. 107745057). D.M.E. wishes to acknowledge and thank the DFG for financial support via DFG individual fellowship number (EV 305/1-1). J.S. acknowledges support from the Alexander von Humboldt Foundation through the Feodor-Lynen fellowship. D.M. thanks NTNU for support through the Onsager Fellowship Program, the Outstanding Academic Fellow Program, and acknowledges funding from the European Research Council (ERC) under the European Union's Horizon 2020 Research and Innovation Program (Grant Agreement No. 863691).